# GRAVITATIONAL POTENTIAL OF CALIFA GALAXIES FROM THE BLUE CLOUD TO THE RED SEQUENCE

VESELINA KALINOVA[1,2]

[1]*Max-Planck-Institut für Radioastronomie (MPIfR), Auf dem Hügel 69, 53121 Bonn, Germany*

[2]*Institute of Astronomy and NAO, Bulgarian Academy of Sciences, 72 Tsarigradsko Shosse, 1784 Sofia, Bulgaria*

E-mail: kalinova@mpifr.de

**Abstract.** We explore the gravitational potential of 238 (E1–Sdm) CALIFA galaxies from the Blue cloud to the Red Sequence using the star formation rate–stellar mass diagram (SFR–$M_\star$) diagnosis. On average, galaxies in the Blue cloud posses slow-rising circular velocity curves (CVCs), those in the Green valley galaxies have flat CVCs, while galaxies in the Red sequence are characterised with peaked CVCs. Both the maximum circular velocity ($V_{\rm c,max}$) and asymptotic circular velocity ($V_{\rm c,flat}$) of the galaxies increase from the Blue cloud to the Red sequence, and show approximately linear relations with the logarithmic offsets from the star-forming main sequence (ΔSFRs). The ΔSFR–$\log_{10}(V_{\rm c,max})$ relation has a steeper slope (–0.154 +/– 0.012) than the ΔSFR–$\log_{10}(V_{\rm c,flat})$ relation (–0.124 +/– 0.011). These results suggest that the $V_{\rm c,max}$, tracing the inner gravitational potential, indicates a stronger variation along the quenching parameter ΔSFR than the $V_{\rm c,flat}$, which is more strongly influenced by the outer mass distribution and dark matter halo.

## 1. INTRODUCTION

The link between the gravitational potential and star-formation activity of galaxies remains a relatively unexplored topic in astrophysics (e.g. Kalinova et al. 2022). The suppression and gradual decline of the star-formation rate (SFR) in galaxies is a complex process known as "star-formation quenching" (e.g. Peng et al. 2010). Understanding the physical mechanisms responsible for quenching and their connection to the mass distribution of galaxies is important for studying galaxy evolution.

An important diagnostic tool for investigating star-formation activity and galaxy quenching is the relation between the SFR and stellar mass ($M_\star$), known as the SFR–$M_\star$ plane (e.g. Brinchmann et al. 2004; Renzini & Peng 2015; González Delgado et al. 2015; Kalinova et al. 2021, 2026). This diagram reveals a well-defined sequence of star-forming galaxies, known as the star-forming main sequence (MS), along which the SFR increases with increasing stellar mass. Galaxies with lower SFRs then the expected ones from this relation populate the region below the MS, including the Green Valley and Red Sequence, and are associated with progressively suppressed star formation. The offset of a galaxy from the MS provides a quantitative measure of its star-formation activity and degree of suppression.

In this work, we calculate the logarithmic offset from the MS (ΔSFR) for each galaxy and investigate its connection with the gravitational potential through circular velocity curve (CVC) properties. This parameter has been widely used in studies of molecular gas properties and star-formation efficiency to investigate the physical processes associated with enhanced or suppressed star formation (e.g. Lin et al. 2019; Ellison et al. 2020; Thorp et al. 2024; Saintonge & Catinella 2022; Colombo et al. 2025a,b). Previous studies have already indicated a connection between star-formation quenching and the gravitational potential and kinematics of galaxies (e.g. Kalinova et al. 2022; Brownson et al. 2022; Kalinova & Colombo 2026). However, here we explore this connection using the offset from the MS as a continuous more general diagnostic tool of global star-formation activity, rather than relying on discrete quenching classifications (Kalinova et al. 2021). We aim to test whether the position of galaxies relative to the MS is reflected in the amplitude and asymptotic velocity, providing further insight into the relation between galaxy dynamics and star-formation quenching.

## 2. DATA & SAMPLE

In this study, we use public available data and catalogues. To study the gravitational potential of the galaxies, we use the CALIFA (Calar Alto Legacy Integral Field spectroscopy Area; Sánchez et al. 2012) CVC catalogue from Kalinova et al. (2017). To derive the CVCs of the galaxies, the authors used Jeans Anisotropic Multi-Gaussian Expansion (JAM; Cappellari 2008, Emsellem et al. 1994) model, assuming constant dynamical mass-to-light ratio and constant velocity anisotropy within 1.5 effective radius of the galaxy. To construct the dynamical models, they also used photometric SDSS DR7 images (Abazajian et al. 2009) and stellar kinematics from CALIFA survey (Falcon-Barroso et al. 2017). More details about the derived dynamical models can be found in Kalinova et al. (2017).

Both $M_\star$ and SFR values of the galaxies comes from Kalinova et al. 2021 derived through spatially resolved stellar-population synthesis with the typical uncertainties adopted from González Delgado et al. (2017) and Cid Fernandes et al. (2014). To obtain the $M_\star$ quantity, the authors fitted the spatially resolved CALIFA COMBO spectra with STARLIGHT (Sánchez et al. 2016; Cid Fernandes et al. 2005; de Amorim et al. 2017), using a Chabrier (2003) IMF. The $M_\star$ from the individual spatial elements were integrated to obtain the total galaxy masses. The SFR was obtained from the reconstructed star-formation histories by summing the $M_\star$ formed during the most recent 32 Myr and dividing by this time interval. This provides a recent, time-averaged SFR and is less dependent on detectable $H_\alpha$ emission than an $H_\alpha$-only estimate.

We exploit the same galaxy sample of 238 (E1–Sdm) as studied in Kalinova et al. (2017). From their work, we also adopt the effective radius ($R_e$) of each system to normalise the radial extent of the galaxy CVCs, allowing comparison on the same scale.

## 3. METHODS

To investigate the variations of the gravitational potential of the galaxies from the Blue cloud to the Red sequence, we use the diagnostic SFR–$M_\star$ diagram and the distance from the star-formation MS for each system. Since the linear model by Elbaz et al. (2007) well describes the distribution of our galaxies in the MS, we adopt this fit throughout our analysis (see solid black line in the left panel of Fig. 1).

The MS model fit is essential for defining the distance of the galaxies from the MS on the SFR–$M_\star$ diagram defined as the logarithmic offset: $\Delta\text{SFR} \equiv \log(\text{SFR}) - \log(\text{SFR}_{\text{MS}})$. Furthermore, we aim to classify the galaxies in three groups: galaxies in the Blue cloud, in the Green valley and in the Red sequence. For the goal, we adapt the definition of Bluck et al. (2016) for defining the Green valley region: $-1.2 < \Delta\text{SFR} < -0.6$. Therefore, the beginning of the Green valley region starts at ΔSFR=−0.6 (indicated by the green dashed line in the left panel of Fig. 1), and ends where ΔSFR=−1.2 (indicated by the red dashed line in the left panel of Fig. 1, marking also the beginning of the Red sequence region).

## 4. RESULTS

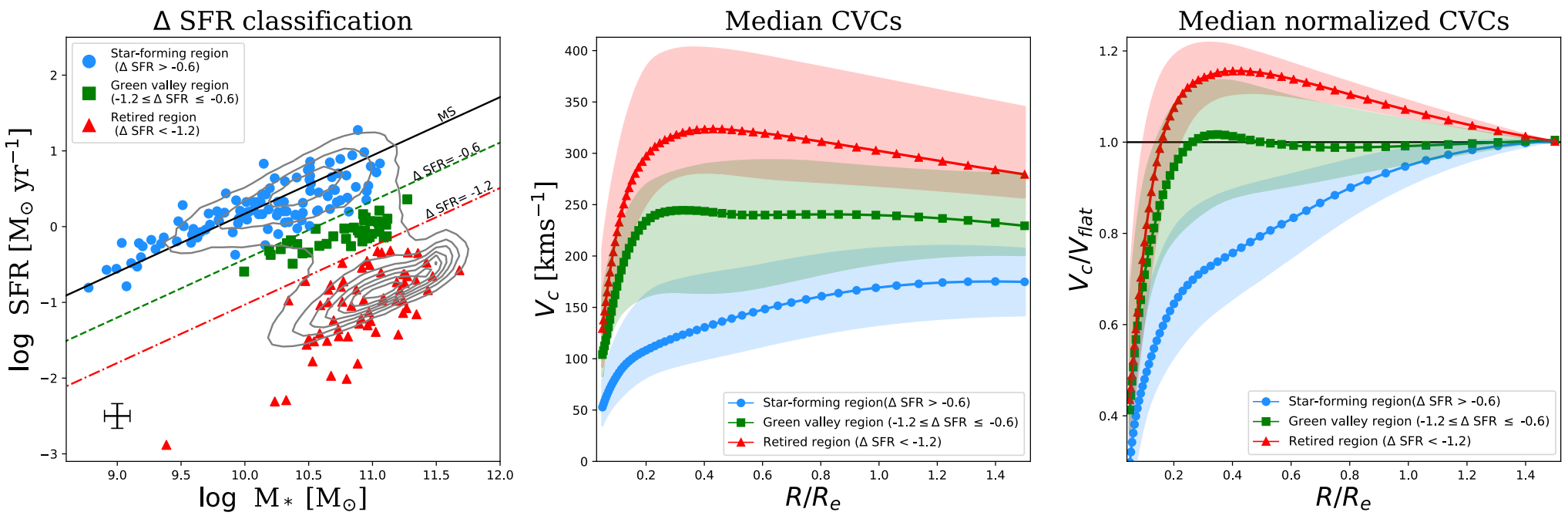

**Figure 1.** *Left:* SFR–$M_\star$ diagram for our sample of 238 CALIFA galaxies with typical uncertainties. The colour-coding of the points indicate the galaxies in the different regions: Blue cloud - blue circle, Green valley - green squares, and Red sequence - red triangles, following the definition of the Green valley from Bluck et al. (2016). The solid black line represents star-forming MS by Elbaz et al. (2007). *Middle*: The median circular velocity curves and their corresponding dispersion bands for each ΔSFR categorisation group using the CALIFA CVC catalog from Kalinova et al. (2017). *Right*: Normalised on the asymptotic velocity median CVC for each ΔSFR group with their corresponding dispersion bands. The horizontal black line indicates the unity of the median velocity and the asymptotic velocity. See Sec. 4 for more details.

Once we obtain the classification label of our galaxies based on their ΔSFR, we separate their individual CVCs in three ΔSFR categories (see top-row panels of Fig. 2), and calculate the corresponding median CVCs for each ΔSFR group (bottom-row panels of Fig. 2). Further, the middle panel of Fig.1 show the combined median CVCs for all three ΔSFR categories, while the right panel of Fig. 1 shows the normalised on the asymptotic velocities median CVCs.

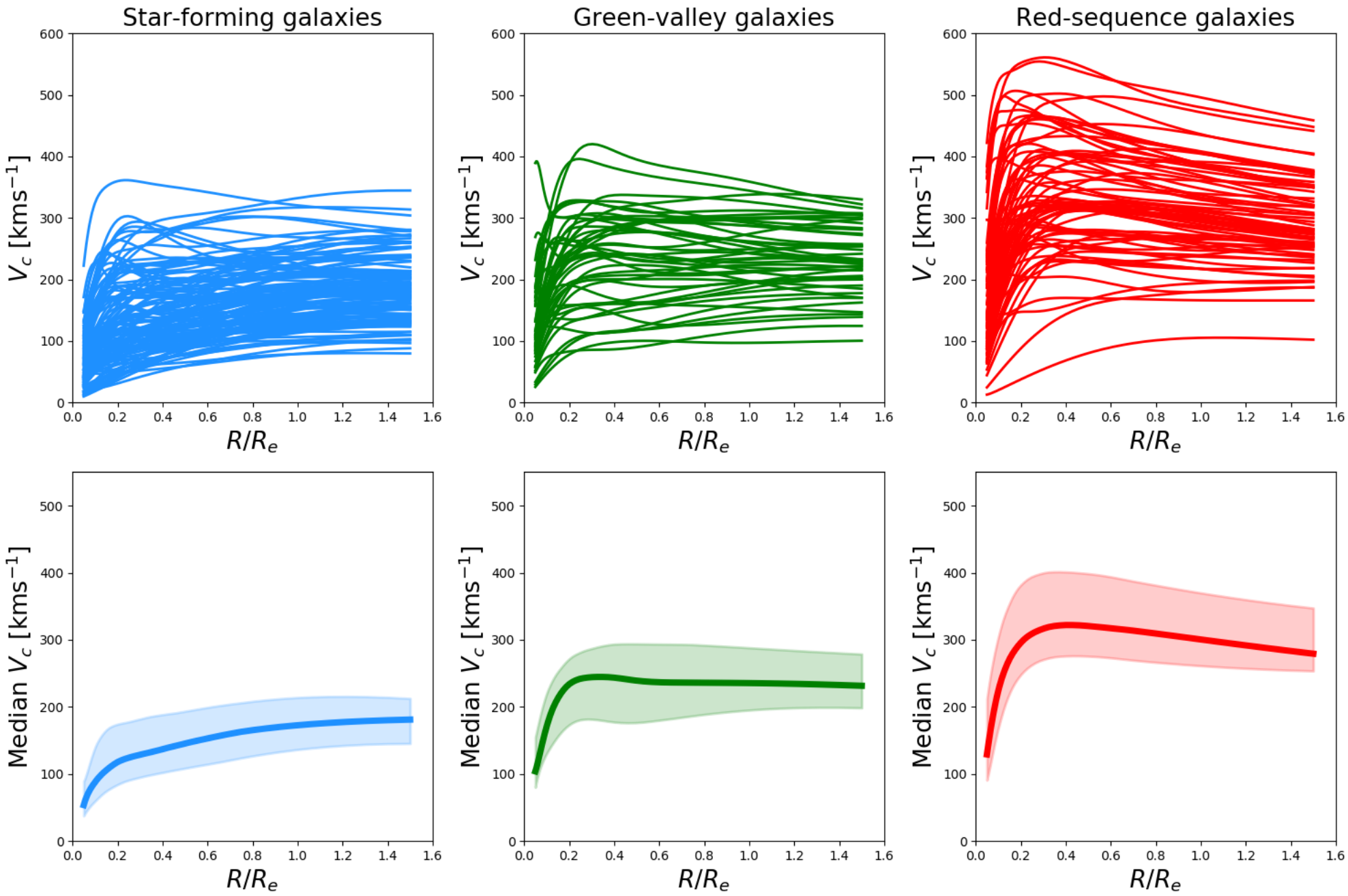

**Figure 2.** Individual circular velocity curves (CVCs) of the studied sample of 238 CALIFA galaxies in each ΔSFR category (top-row panels), and the corresponding median CVCs of the individual curves for each ΔSFR group (bottom-row panels). The colour-coding distinguishes star-forming galaxies in the Blue Cloud (blue stars), Green Valley galaxies (green triangles), and Red Sequence galaxies (red circles). See Sec. 4 for more details.

Given the results in Fig. 1 and Fig. 2, we observe that galaxies have a broad span in CVC amplitudes and shapes from the Blue cloud to the Red sequence. However, once combined for each ΔSFR group and calculated their prototype (median) curve, they show very specific shape and amplitude. For example, Blue cloud group is characterised with slow-rising prototype CVC within velocity range of 40–100 km/s for the inner part and 120–180 km/s to the outer part. The Green valley region is characterised with flat CVC within velocity range of 100–270 km/s for the inner part and 180–270 km/s for the outer part. The Red sequence group are characterised by prototype curve with prominent peak with velocity range of 150–400 km/s for the inner and 270–340 km/s for the outer part.

Overall, from Fig. 1 we observe the building up of the central parts of the galaxies from the Blue cloud to the Red sequence (right panel), as well as growing up of the galaxy disk through increasing of the overall asymptotic circular velocity across ΔSFR group (middle panel). Similar smooth transitions in the gravitational potentials of the galaxies and their CVCs across the SFR–$M_\star$ diagram have been reported in previous studies, but using a different categorisation scheme (Kalinova et al. 2021, 2022, 2026).

To better understand the link between the quenching parameter ΔSFR and the properties of the gravitational potential of galaxies, we investigate the relations between the CVC amplitude ($V_{c,max}$) of each galaxy and their corresponding ΔSFR, as well as the asymptotic circular velocity ($V_{c,flat}$) and the ΔSFR (see Fig. 3). We observe linear dependencies of both $V_{c,max}$ and $V_{c,flat}$ on ΔSFR, with few outliers. We fit both relations using the robust Theil–Sen estimator (Theil 1950; Sen 1968), which determines the slope as the median of all pairwise slopes and reduces the influence of outliers. Both relations exhibit negative slopes, indicating a systematic increase in circular velocity towards galaxies progressively below the star-formation MS. The relation based on $V_{c,max}$ shows a steeper decline (with approximately 24 %) than the one based on $V_{c,flat}$ (see also Table 1). This difference suggests that the $V_{c,max}$, which is more sensitive to the inner mass distribution, indicates a stronger dependence on the quenching parameter ΔSFR than the asymptotic velocity. The uncertainties of the Theil–Sen linear fits are estimated using bootstrap resampling with 1000 iterations.

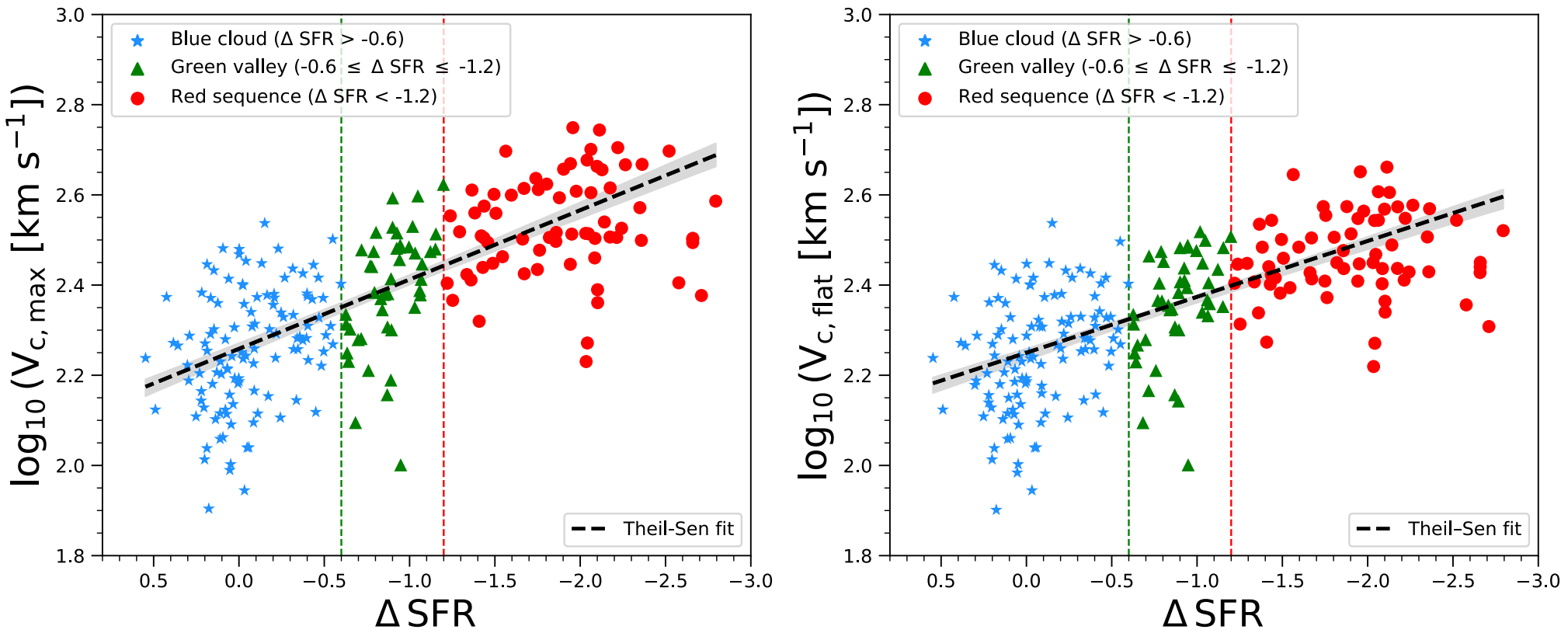

**Figure 3.** Circular velocity amplitude ($V_{c,max}$; left) and asymptotic circular velocity ($V_{c,flat}$; right) of individual galaxies, from the Blue cloud to the Red sequence. The dashed black lines represent the Theil–Sen linear fits to the relations, while the shaded regions indicate the 1σ uncertainty ranges of the fits. The green and red dashed lines indicate the region of the Green valley. The colour–coding marks the galaxies in the different quenching regions on the SFR–$M_\star$ diagram: Blue cloud (blue stars), Green valley (green triangles), and Red sequence (red circles). See Sec. 4 for more details.

| *Relation* | *a (intercept)* | *b (slope)* |
|---|---|---|
| ΔSFR–$\log_{10}(V_{c,max})$ | 2.259 +/–0.014 | -0.154 +/–0.012 |
| ΔSFR–$\log_{10}(V_{c,flat})$ | 2.250 +/– 0.012 | -0.124 +/– 0.011 |

**Table 1.** Theil-Sen linear fits to the of ΔSFR-velocity relations for our sample of 238 CALIFA nearby galaxies. See Sec. 4 for more details.

## 5. SUMMARY & CONCLUSIONS

We analysed the gravitational potential of 238 E1–Sdm nearby CALIFA galaxies through their CVCs across the SFR–$M_\star$ diagram. On average, galaxies in the Blue cloud posses slow-rising CVCs, the Green valley galaxies have flat CVCs, and the Red sequence galaxies are characterised with peaked CVCs. The amplitudes and asymptotic circular velocities gradually increase from the Blue cloud to the Red sequence.

Furthermore, we investigate the link between the ΔSFR and the CVCs for each galaxy. We found that that both $V_{c,max}$ and $V_{c,flat}$ increase towards increasingly negative ΔSFR, indicating higher characteristic circular velocities of the galaxies progressively below the star-formation MS. The ΔSFR–$\log_{10}(V_{c,max})$ relation has a steeper slope (–0.154 +/– 0.012) than the ΔSFR–$\log_{10}(V_{c,flat})$ relation (–0.124 +/– 0.011). Outlier-resistant Theil-Sen models are fitted to both ΔSFR-velocity relations with uncertainties estimated through 1000 bootstrap iterations.

These approximately linear relations suggest that the kinematic/dynamical properties of the galaxies vary progressively with star-formation activity. This also indicates a relatively stronger dependency of the inner gravitational potential of the galaxies, mainly influenced by the bulge/spheroid of the galaxies, on the quenching parameter ΔSFR in comparison to the outer gravitational potential of the galaxies, which is mainly controlled by the dark matter halo. It seems that the inner and outer gravitational potentials respond differently to the processes associated with star-formation quenching.

## Acknowledgements

In this study, we made use of the data of the first legacy survey, the Calar Alto Legacy Integral Field Area (CALIFA) survey, based on observations made at the Centro Astronómico Hispano Alemán (CAHA) at Calar Alto, operated jointly by the Max Planck-Institut für Astronomie and the Instituto de Astrofísica de Andalucía (CSIC). This research made use of the open-source python packages as Astropy (Price-Whelan et al., 2018),

SciPy (Virtanen et al., 2020), NumPy (Harris et al., 2020), and Matplotlib (Hunter, 2007). V. K. thanks Dario Colombo and Keiichi Kodaira for the useful discussions.

## References

Abazajian, K. N., Adelman-McCarthy, J. K., Agüeros, M. A., et al.: 2009, *ApJS*, **182**, 543–558.
Bluck, A. F. L., Mendel, J. T., Ellison, S. L., et al.: 2016, *MNRAS*, **462**, 2559–2586.
Brinchmann, J., Charlot, S., White, S. D. M., et al.: 2004, *MNRAS*, **351**, 1151–1179.
Brownson, S., Bluck, A. F. L., Maiolino, R., & Jones, G. C.: 2022, *MNRAS*, **511**, 1913–1941.
Cappellari, M.: 2008, *MNRAS*, **390**, 71–86.
Chabrier, G.: 2003, *PASP*, **115**, 763–795.
Cid Fernandes, R., Mateus, A., Sodré, L., Stasińska, G., & Gomes, J. M.: 2005, *MNRAS*, **358**, 363–378.
Cid Fernandes, R., González Delgado, R. M., García Benito, R., et al.: 2014, *A&A*, **561**, A130.
Colombo, D., Kalinova, V., Bazzi, Z., et al.: 2025a, *A&A*, **699**, A366.
Colombo, D., Kalinova, V., Bazzi, Z., et al.: 2025b, *A&A*, **699**, A367.
de Amorim, A. L., et al.: 2017, *MNRAS*, **471**, 3727–3752.
Elbaz, D., Daddi, E., Le Borgne, D., et al.: 2007, *A&A*, **468**, 33–48.
Ellison, S. L., Thorp, M. D., Lin, L., et al.: 2020, *MNRAS*, **493**, L39–L43.
Emsellem, E., Monnet, G., & Bacon, R.: 1994, *A&A*, **285**, 723–738.
Falcón-Barroso, J., Lyubenova, M., van de Ven, G., et al.: 2017, *A&A*, **597**, A48.
González Delgado, R. M., García-Benito, R., Pérez, E., et al.: 2015, *A&A*, **581**, A103.
González Delgado, R. M., Pérez, E., Cid Fernandes, R., et al.: 2017, *A&A*, **607**, A128.
Harris, C. R., Millman, K. J., van der Walt, S. J., et al.: 2020, *Nature*, **585**, 357–362.
Hunter, J. D.: 2007, *Computing in science & engineering*, **9**, 90–95.
Kalinova, V., & Colombo, D.: 2026, *J. Phys.: Conf. Ser.*, **3282**, 012003.
Kalinova, V., Colombo, D., Rosolowsky, E., et al.: 2017, *MNRAS*, **469**, 2539–2594.
Kalinova, V., Colombo, D., Sánchez, S. F., et al.: 2021, A&A, **648**, A64.
Kalinova, V., Colombo, D., Sánchez, S. F., et al.: 2022, A&A, **665**, A90.
Kalinova, V., Kodaira, K., & Colombo, D.: 2026, *Bulgarian Astronomical Journal*, 45, 54–73.
Lin, L., Pan, H.-A., Ellison, S. L., et al.: 2019, *ApJL*, **884**, L33.
Peng, Y.-j., Lilly, S. J., Kovač, K., et al.: 2010, ApJ, **721**, 193–221.
Price-Whelan, A. M., Sipőcz, B., Günther, H., et al.: 2018, *The Astronomical Journal*, **156**, 123.
Renzini, A., & Peng, Y.-j.: 2015, *ApJ*, **801**, L29.
Saintonge, A., & Catinella, B.: 2022, *ARA&A*, **60**, 319–361.
Sánchez, S. F., Kennicutt, R. C., Gil de Paz, A., et al.: 2012, *A&A*, **538**, A8.
Sánchez, S. F., et al.: 2016, *A&A*, **594**, A36.
Sen, P. K.: 1968, *Journal of the American Statistical Association*, **63**, 1379–1389.
Theil, H.: 1950a, *Proc. Kon. Ned. Akad. Wetensch.*, **53**, 386–392.
Thorp, M. D., Ellison, S. L., & Galicia, A.: 2024, ***A&A***, 690, L4.
Virtanen, P., Gommers, R., Oliphant, T. E., et al., and SciPy 1.0 Contributors: 2020, *Nature Methods*, **17**, 261–272.